\documentclass[11pt]{article}
\usepackage{graphicx}
\usepackage{amssymb}
\usepackage{epstopdf}
\usepackage{epsfig}
\usepackage{amsmath,amssymb,mathrsfs}
\usepackage[latin1]{inputenc}
\usepackage{graphicx}
\newcommand{}{{{\^\i}}}
\newcommand{}{{\'{e}}}
\newcommand{}{{\`{e}}}
\newcommand{}{{\^{e}}}
\newcommand{}{{\"{e}}}
\newcommand{}{{\`{a}}}
\newcommand{}{{{a}}}
\newcommand{}{{\"{a}}}
\newcommand{}{{\`{u}}}
\newcommand{}{{{u}}}
\newcommand{}{{\"{u}}}
\newcommand{}{{\`{o}}}
\newcommand{}{{{o}}}
\newcommand{}{{\"{o}}}

\title{Information Theory, Relative Entropy and Statistics}
\author{Fran\c{c}ois~Bavaud\\ University of Lausanne\\
  Switzerland\\}

\date{This article has been published as: \\   {\small Bavaud F. (2009) {\em Information Theory, Relative Entropy and Statistics}. \\ In: Sommaruga G. (editor): {\em Formal Theories of Information}. Lecture Notes in Computer Science 5363, Springer, Berlin, pp. 54--78.}}
 
\begin{document}
\maketitle

\section{\large Introduction:  the relative entropy as an epistemological functional}
Shannon's Information Theory (IT) (1948) definitely established the purely {\em mathematical} nature of 
 entropy and relative entropy, in contrast to the previous identification by Boltzmann (1872) of his ``$H$-functional" as the {\em physical} entropy of earlier thermodynamicians (Carnot, Clausius, Kelvin).  The following declaration is attributed to Shannon  (Tribus and McIrvine 1971): 
 
  \begin{center}\begin{minipage}{12.2cm}
 {\em My greatest concern was what to call it. I thought of calling it ``information", but the word
was overly used, so I decided to call it ``uncertainty". When I discussed it with John von
Neumann, he had a better idea. Von Neumann told me, ``You should call it entropy, for two
reasons. In the first place your uncertainty function has been used in statistical mechanics
under that name, so it already has a name. In the second place, and more important, nobody
knows what entropy really is, so in a debate you will always have the advantage."}
\end{minipage}\end{center}

 In IT, the entropy of a message limits its minimum coding length, 
 in the same way that, more generally, the complexity of the message determines its compressibility  in the Kolmogorov-Chaitin-Solomonov algorithmic information theory (see e.g. Li and Vitanyi (1997)). 
 
Besides coding and compressibility interpretations, the relative entropy also turns out to possess a direct probabilistic meaning, as demonstrated by the {\em asymptotic rate formula} (\ref{typesform}). 
This circumstance enables a complete exposition of  classical inferential statistics (hypothesis testing, maximum likelihood, maximum entropy, exponential and log-linear models, EM algorithm, etc.) under the guise of a discussion of the properties of the relative entropy.

 In a nutshell, the relative entropy $K(f||g)$ has two arguments $f$ and $g$, which both are probability distributions belonging to the same simplex. Despite formally similar, the arguments are epistemologically contrasted:  $f$ represents  the observations, the data,   what {\em we see}, while $g$ represents the expectations, the models, what  {\em we believe}.   $K(f||g)$ is an asymmetrical measure of dissimilarity between empirical and  theoretical distributions, able to capture
  the various aspects of the confrontation between   models and data, that  is  the art of classical statistical inference, including 
 Popper's refutationism as a particulary case. Here lies the dialectic charm of $K(f||g)$, which emerges in that respect as an  {\em epistemological functional}.

We have here attempted to emphasize and synthetize the conceptual significance of the theory, rather than insisting on its mathematical rigor, the latter being  thoroughly developped in a broad and widely available litterature (see e.g. Cover and Thomas (1991) and references therein). Most of the illustrations bear on independent and identically distributed (i.i.d.) finitely valued observations, that is on {\em dice models}.  This convenient restriction is not really limiting, and can be extended to Markov chains of finite order, as  illustrated in the last part on textual data with presumably original  applications, such as heating and cooling texts, or additive and multiplicative text mixtures.

\section{The asymptotic rate formula}
\subsection{Model and empirical distributions}
 $D=(x_1 x_2\ldots x_n)$  denotes  the data,  consisting of $n$ observations, and
$M$ denotes a possible model for those data. The corresponding probability is $P(D|M)$, with 
\begin{eqnarray*}
 P(D|M)\ge0\qquad\qquad\qquad \sum_D P(D|M)=1.
\end{eqnarray*}
 Assume (dice models) that 
each observation can take on $m$ discrete values, each observation $X$ being 
 i.i.d. distributed as 
\begin{eqnarray*}
f_j^M:=P(X=j)\qquad\qquad j=1,\ldots,m.
\end{eqnarray*}
$f^M$ is the {\em model  distribution}. The {\em empirical distribution}, also called {\em type}    
(Csisz{\'a}r and K\"{o}rner 1980) in the IT framework,  is \begin{eqnarray*}
f_j^D:=\frac{n_j}{n}\qquad\qquad j=1,\ldots,m
\end{eqnarray*}
where $n_j$ counts the  occurences of the $j$-th category and  $n=\sum_{j=1}^m n_j$ is the sample size.

\begin{figure}
    \begin{center}
\includegraphics[width=2.5in]{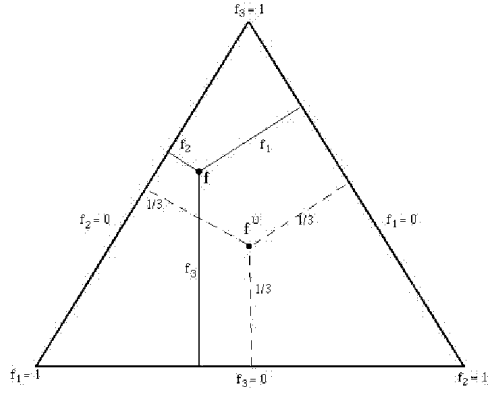}
  \end{center}
 \caption{The simplex $S_3$, where $f^U=(\frac13,\frac13,\frac13)$ denotes the uniform distribution. In the interior of $S_m$, a distribution $f$ can be varied along
  $m-1$ 
 independant directions, that is $\mbox{\small dim}(S_m)=m-1$.} \label{simplex}\end{figure}

 Both $f^M$ and $f^D$  are discrete distributions  with $m$ modalities. Their collection form the {\em simplex} $S_m$ (figure \ref{simplex})
 \begin{eqnarray*}
S\equiv S_m:=\{f\: |\: f_j\ge0 \quad \mbox{\small and}\quad \sum_{j=1}^mf_j=1\}.\end{eqnarray*}
  \subsection{Entropy and relative entropy: definitions and properties}
 Let $f,g\in S_m$. The {\em entropy} $H(f)$ of $f$ and the {\em relative entropy} $K(f||g)$ between $f$ and $g$ are defined (in   nats) as
\begin{eqnarray*}
H(f) & := &-\sum_{j=1}^m f_j \ln f_j
 =  \mbox{\small entropy of $f$}
\\
K(f||g) & := &\sum_{j=1}^m f_j \ln \frac{f_j}{g_j}
 =  \mbox{\small relative entropy of $f$ with respect to $g$}\: .
\end{eqnarray*}
 $H(f)$ is concave in $f$, and constitutes a measure of the {\em uncertainty of the outcome}  among $m$ possible outcomes (proofs are standard):
\begin{eqnarray*}
0\le  H(f) \le \ln m
\end{eqnarray*}
where 
\begin{itemize}
  \item[$\bullet$] $H(f)=0$ iff $f$ is a {\em deterministic} distribution concentrated on a single modality (minimum  uncertainty)
 \item[$\bullet$] $H(f)=\ln m $ iff $f$ is the {\em uniform} distribution (of the form  $f_j=1/m$) (maximum  uncertainty).
 \end{itemize}
 
 \vspace{0.1in}

 $K(f||g)$, also known as  the {\em Kullback-Leibler divergence},  is convex in both arguments, and constitutes a {\em non-symmetric} measure of the {\em dissimilarity} between the distributions $f$ and $g$, with 
\begin{eqnarray*}
0\le  K(f||g) \le \infty
\end{eqnarray*}
where 
\begin{itemize}
  \item[$\bullet$] $K(f||g)=0$ iff $f\equiv g$  
 \item[$\bullet$] $K(f||g)<\infty $ iff   $f$ is absolutely continuous with respect to $g$, that 
 is if $g_j=0$ implies $f_j=0$.
  \end{itemize}
  
  \vspace{0.1in}

  Let the  categories $j=1,\ldots, m$ be {\em coarse-grained}, that is aggregated into groups of 
  super-categories
$J=1,\ldots, M<m$. Define
\begin{eqnarray*}
F_J:=\sum_{j\in J}f_j\qquad\qquad G_J:=\sum_{j\in J}g_j\:. \end{eqnarray*}
Then 
\begin{eqnarray}
H(F)\le H(f) \qquad\qquad K(F||G)\le K(f||g)\: .
\label{corgr}
\end{eqnarray}

\subsection{Derivation of the asymptotic rate (i.i.d. models)} 
\label{derivtyp}
On one hand, straightforward algebra yields
 \begin{eqnarray}
\label{type1}
P(D|f^M):=P(D|M)&=&P(x_1 x_2\ldots x_n|M)=\prod_{i=1}^n (f^M_j)^{n_j}\notag \\
& =&
\exp[-n K(f^D||f^M) - nH(f^D)]\enspace .
\end{eqnarray}
On the other hand, each permutation of the data $D=(x_1,\ldots, x_n)$ yields the same 
 $f^D$.
 Stirling's approximation $n!\cong n^n \exp(-n)$ (where $a_n\cong b_n$ means $\lim_{n\to \infty}\frac1n\ln(a_n/b_n)=0$) shows that 
 \begin{equation}
\label{type3}
P(f^D|M)=\frac{n!}{n_1!\cdots n_m!}\: P(D|M) \cong
\exp(n H(f^D))\: P(D|M).
\end{equation}
(\ref{type1}) and (\ref{type3}) imply the {\bf asymptotic rate formula}:
\begin{eqnarray}
 \label{typesform}
 P(f^D|f^M) \cong
\exp(- n \: K(f^D||f^M))\qquad \qquad \qquad  \mbox{\small asymptotic rate formula}\:.
\end{eqnarray}
Hence, $K(f^D||f^M)$ is the  asymptotic  rate of the quantity $P(f^D|f^M)$,  the {\em probability} of the empirical distribution $f^D$ for a given model $f^M$, or equivalently
the {\em likelihood} of the model $f^M$ for the data  $f^D$. Without additional constraints, the model $\hat{f}^M$ maximizing the likelihood is simply $\hat{f}^M=f^D$ (section \ref{mlht}). Also, without further information, the most probable empirical distribution $\tilde{f}^D$ is simply $\tilde{f}^D=f^M$
 (section \ref{maoIAUS}).

\subsection{Asymmetry of the relative entropy and hard falsificationism} 
$K(f||g)$ as a  dissimilarity measure between  $f$ and $g$ is {\em proper} (that is $K(f||g)=0$ implies $ f\equiv g$) but {\em not  symmetric}
($K(f||g) \neq K(g||f)$ in  general). Symmetrized dissimilarities  such as $J(f||g):=\frac12(K(f||g)+K(g||f))$ or $L(f||g):=K(f||\frac12(f+g))+K(g||\frac12(f+g))$ have often been proposed in the literature. 

\vspace{0.1in}

The conceptual significance of such functionals can indeed be questioned: from equation 
 (\ref{typesform}), the first argument $f$ of $K(f||g)$ should be an empirical distribution, and the second argument $g$ a model distribution. 
Furthermore,  {\em the asymmetry of the relative entropy does  not  constitute a defect, but  perfectly matches the asymmetry between data and models}. Indeed

\begin{itemize}
  \item[$\bullet$] if $f_j^M=0$ and $f_j^D>0$, then  $K(f^D||f^M)=\infty$ and, from  (\ref{typesform}), $ P(f^D|f^M)=0$ and, unless the veracity of the data $f^D$ is questioned, the model distribution $f^M$ should be strictly rejected
 \item[$\bullet$] if on the contrary $f_j^M>0$ and $f_j^D=0$, then  $K(f^D||f^M)< \infty$ and $ P(f^D|f^M)>0$ in general,  and $f^M$ should not be rejected, at least for small samples. 
  \end{itemize}

Thus the theory  ``All crows are black" is refuted by the single observation of a white crow, while the theory   ``Some crows are black" is not refuted by the observation of a thousand white crows. In this spirit, 
Popper's  falsificationist mechanisms (Popper 1963) are captured by the properties of the relative entropy, and can be further extended to  probabilistic or ``soft falsificationist" situations, beyond the purely logical true/false context   (see section \ref{fiuhui}).

\subsection{The chi-square approximation}
Most of the properties of the relative entropy  are shared by another 
functional, historically  anterior and  well-known to statisticians,  namely the {\em chi-square} $\chi^2(f||g):=n\sum_j (f_j-g_j)^2/g_j$. As a matter of fact, the relative entropy and the chi-square (divided by $2n$) are identical up to the third order: 
   \begin{equation}
  \label{chsqsqsqs}
\mbox{\small $2 K(f||g)=\sum_{j=1}^m \frac{(f_j-g_j)^2}{g_j}+O(\sum_j \frac{(f_j-g_j)^3}{g_j^2})=\frac{1}{n}\chi^2(f||g)+O(||f-g||^3)$}
\end{equation}

\subsubsection{Example: coin ($m=2$)}
\label{qqq34}
The values of the relative entropy and the chi-square read, for various $f^M$  and $f^D$, as : 

\begin{center}\begin{tabular}{|c|c|c|c|c|}
\hline
 & $f^M$ & $f^D$ & $K(f^D||f^M)$ & $\chi^2(f^D||f^M)/2n$ \\ \hline
a)    & $(0.5,0.5)$ & $(0.5,0.5)$ & $0$ & $0$  \\
b)    & $(0.5,0.5)$ & $(0.7,0.3)$ & $0.0823$ & $0.08$  \\
c)    & $(0.7,0.3)$ & $(0.5,0.5)$ & $0.0822$ & $0.095$  \\
d)    & $(0.7,0.3)$ & $(0.7,0.3)$ & $0$ & $0$  \\
e)    & $(0.5,0.5)$ & $(1,0)$ & $0.69$ & $0.5$  \\
f)    & $(1,0)$ & $(0.99,0.01)$ & $\infty$ & $\infty$  \\
\hline
\end{tabular}\end{center}

 \section{Maximum likelihood and hypothesis testing}
 \label{mlht}
 \subsection{Testing a single hypothesis (Fisher)}
 \label{fiuhui}
As shown by (\ref{typesform}), the  higher $K(f^D||f^M)$, the lower the likelihood $P(f^D|f^M)$. This circumstance permits to test the single hypothesis $H_0$ :  ``{\em the model distribution is $f^M$}". If $H_0$ were true, $f^D$ should fluctuate around its expected value $f^M$, and fluctuations of too large amplitude, with occurrence probability  less than $\alpha$ (the significance level), should lead to the rejection of $f^M$.  Well-known  results
on the chi-square distribution  (see e.g. Cramer (1946) or Saporta (1990)) together with  approximation (\ref{chsqsqsqs}) shows  $2n K(f^D||f^M)$ to be distributed,  under $H_0$ and  for
 $n$ large, as $\chi^2[{\scriptstyle{\mbox{df}}}]$ with $\mbox{df}=\mbox{\small dim}(S_m)=m-1$ degrees of freedom. 

\vspace{0.1in}

Therefore, the test consists in {\em  rejecting $H_0$ at level $\alpha$} if 
\begin{eqnarray}
\label{testsimp}
2\: n\: K(f^D||f^M)\ge \chi^2_{1-\alpha}\scriptstyle{[m-1]}\: .
\end{eqnarray}
In that respect, Fisher's classical hypothesis testing appears as a {\em soft falsificationist} strategy, yielding the rejection of a theory $f^M$ for large values of $ K(f^D||f^M)$. It  generalizes
  Popper's (hard) falsificationism which is limited to situations of strict refutation as expressed by $K(f^D||f^M)=\infty$.

\subsection{Testing a family of models}
\label{famtest}
Very often, the hypothesis to be tested is {\em composite}, that is of the form $H_0$ :  ``{\em $f^M\in {\cal M}$}", where ${\cal M}\subset S=S_m$ constitutes a family of models containing a number $\mbox{\small dim}({\cal M})$  of free, non-redundant parameters.

If the observed distribution itself satisfies $f^D\in {\cal M}$, then there is obviously no reason to reject 
$H_0$. But $f^D\notin {\cal M}$ in general, and hence
\begin{eqnarray*}
\mbox{\small $\min_{f\in{\cal M}}K(f^D||f)=K(f^D||\hat{f}^{\cal{M}})$}
\:\mbox{\small is strictly positive, with}\:
\mbox{\small $\hat{f}^{\cal{M}}:=\arg\min_{f\in {\cal{M}}}\: K(f^D||f)\:. $}
\end{eqnarray*}
$\hat{f}^{\cal{M}}$ is known as the  {\em maximum likelihood estimate} of the model, and depends on both $f^D$ and ${\cal M}$. We assume $\hat{f}^{\cal{M}}$ to be unique, which is e.g.  the case if  
${\cal M}$ is  convex. 

 \vspace{0.1in}
 
If $f^M\in {\cal M}$, $2nK(f^D||\hat{f}^{\cal{M}})$ follows a chi-square distribution with 
$\mbox{\small dim}({\cal S})-\mbox{\small dim}({\cal M})$ degrees of freedom. Hence, 
one   rejects $H_0$   at level $\alpha$ if 
\begin{eqnarray}
\label{testcomp}
2n K(f^D||\hat{f}^{\cal{M}})\ge \chi^2_{1-\alpha}\scriptstyle{[\dim(S)-\dim(\cal{M})]}\: .
\end{eqnarray}
If ${\cal M}$ reduces to a unique distribution $f^M$, then $\dim({\cal M})=0$ and (\ref{testcomp}) reduces to (\ref{testsimp}). In the opposite direction,   ${\cal M}=S$ defines the {\em saturated} model, in which case  (\ref{testcomp}) yields the undefined inequality $0\ge  \chi^2_{1-\alpha}[0]$.

\subsubsection{Example: coarse grained model specifications}
\label{qqq087}
Let $f^M$ be a dice model, with categories $j=1,\ldots,m$. Let $J=1,\ldots,M<m$ denote {\em groups of categories}, and suppose that the model specifications are coarse-grained  (see (\ref{corgr})), that is 
 \begin{eqnarray*}
{\cal M}=\{\: f^M\:  |\: \sum_{j\in J}f_j^M\stackrel{!}{=}F_J^M\quad J=1,\ldots,M\: \}\: .
\end{eqnarray*}
Let $J(j)$ denote the group to which $j$ belongs. Then the maximum likelihood (ML) estimate is simply
\begin{eqnarray}
\label{cocococ}
 \hat{f}^{\cal M}_j=f_j^D   
  \: \frac{F_{J(j)}^M}{F_{J(j)}^D}\: \: \: \: \: \: \: \mbox{\small where}\: \:  F_J^D:=\sum_{j\in J}f_j^D\: \mbox{\small and}\: \:  K(f^D||\hat{f}^{\cal{M}})=K(F^D||F^M).   \end{eqnarray}

 \subsubsection{Example: independence}
\label{qqq088}
  Let $X$ and $Y$ two categorical variables  with  modalities $j=1,\ldots,m_1$ and $k=1,\ldots,m_2$. Let   $f_{jk}$ denote the joint distribution of $(X,Y)$. The distribution of $X$ alone (respectively $Y$ alone) obtains as the marginal $f_{j\bullet}:=\sum_k f_{jk}$ (respectively $f_{\bullet k}:=\sum_j f_{jk}$). Let ${\cal M}$ denote the set of  {\em independent distributions}, i.e. 
\begin{eqnarray*}
{\cal M}=\{ f\in S\: |\: f_{jk}=a_jb_k \}\: .
\end{eqnarray*}
The corresponding ML estimate $\hat{f}^M\in {\cal M}$ is
\begin{eqnarray*}
\hat{f}^{\cal M}_{jk}=f^D_{j\bullet}\: f^D_{ \bullet k}\qquad\mbox{\small where}\quad
f^D_{j\bullet}:=\sum_k f_{jk}^D\quad \mbox{\small and}\quad f^D_{\bullet k}:=\sum_j f_{jk}^D
\end{eqnarray*}
 with the well-known property (where $H_D(.)$ denotes  the entropy associated to the empirical distribution)
 \begin{eqnarray}
 \label{indeptrio}
\mbox{\small  $K(f^D||\hat{f}^{\cal M})=H_D(X)+H_D(Y)-H_D(X,Y)= \frac12\sum_{jk}\frac{(f_{jk}^D-\hat{f}_{jk}^{\cal M})^2}{\hat{f}_{jk}^{\cal M}}+0(||f^D-\hat{f}^{\cal M}||^3)$}\: .
\end{eqnarray}
The  {\em mutual information} $I(X:Y):=H_D(X)+H_D(Y)-H_D(X,Y)$  is the information-theoretical measure  of  dependence between $X$ and $Y$. Inequality $H_D(X,Y)\le H_D(X)+H_D(Y)$ insures its non-negativity. By (\ref{indeptrio}), the corresponding test reduces to the usual   chi-square test of independence, with $\dim(S)-\dim({\cal{M}})=
(m_1m_2-1)-(m_1+m_2-2)=(m_1-1)(m_2-1)$ degrees of freedom.

  \subsection{Testing between two hypotheses (Neyman-Pearson)}
Consider the two hypotheses $H_0$ : `` {\em $f^M=f^0$} " and $H_1$ : `` {\em $f^M=f^1$} ", where 
$f^0$ and  $f^1$ constitute two distinct  distributions in $S$. Let $W\subset S$ denote the {\em rejection region} for $f^0$, that is such that $H_1$ is accepted if $f^D\in W$, and $H_0$ is accepted if $f^D\in W^c:=S\setminus W$. 
The errors of first, respectively second kind are
\begin{eqnarray*}
\alpha:=P(f^D\in W\: |\: f^0)\qquad\qquad
\beta:=P(f^D\in W^c\: |\: f^1)\: .\label{albet}
\end{eqnarray*}
For $n$ large, Sanov's theorem (\ref{maxent00}) below shows that 
\begin{eqnarray}\label{alplaoj}
\alpha\cong \exp(-n K(\tilde{f}^0||f^0))\qquad  
\tilde{f}^0:=\arg\min_{f\in W} K(f||f^0)\\
\beta\cong \exp(-n K(\tilde{f}^1||f^1))\qquad 
\tilde{f}^1:=\arg\min_{f\in W^c} K(f||f^1) . \notag
\end{eqnarray}

\vspace{0.1in}

The rejection region $W$ is said to be {\em optimal} if there is no other region $W'\subset S$ with $\alpha(W')<\alpha(W)$ and $\beta(W')<\beta(W)$. The celebrated Neyman-Pearson lemma, together with the asymptotic rate formula (\ref{typesform}),
 states that $W$ is optimal iff it is of the form 
\begin{eqnarray}
\label{npoptrr}
W=\{f \: |\: \frac{P(f |f^1)}{P(f |f^0)}\ge T\}=\{f \: |\: K(f||f^0)-K(f||f^1)\ge \frac{1}{n}\ln T:=\tau\}
\end{eqnarray}

One can demonstrate (see e.g. Cover and Thomas (1991) p.309) that the distributions (\ref{alplaoj}) governing the asymptotic error rates coincide when $W$ is optimal, and are given by the {\em multiplicative mixture}
\begin{eqnarray}
\label{neopea}
\tilde{f}^0_j=\tilde{f}^1_j=f_j(\mu):=\frac{(f^0_j)^\mu (f^1_j)^{1-\mu}}{\sum_k (f^0_k)^\mu (f^1_k)^{1-\mu}}
\end{eqnarray}
where $\mu$ is the value insuring $K(f(\mu)||f^0)-K(f(\mu)||f^1)=\tau$. Finally, the overall probability of error, that is the probability of occurrence of an error of first {\em or} second kind, is minimum for $\tau=0$, with rate  equal to 
\begin{eqnarray*}
K(f(\mu^\ast)||f^0)=K(f(\mu^\ast)||f^1)=-\min_{0\le\mu\le 1}\: \ln(\sum_k (f^0_k)^\mu (f^1_k)^{1-\mu})=:C(f^0,f^1)
\end{eqnarray*}
where $\mu^\ast$ is the value minimising the third term. The quantity $C(f^0,f^1)\ge0$, known as {\em Chernoff information}, constitutes a symmetric dissimilarity between the distributions $f^0$ and $f^1$, and measures how easily $f^0$ and $f^1$ can be discriminated from each other. In particular, $C(f^0,f^1)=0$ iff $f^0=f^1$. 

\subsubsection*{Example \ref{qqq34}, continued: coins}
Let $f:=(0.5,0.5)$, $g:=(0.7,0.3)$, $h:=(0.9,0.1)$ and $r:=(1,0)$. Numerical estimates yield (in  nats) $C(f,g)=0.02$, $C(f,h)=0.11$, 
$C(g,h)=0.03$ and $C(f,r)=\ln 2=0.69$.

 \subsection{Testing a family within another}
 \label{tetene}
 Let ${\cal M}_0$ and ${\cal M}_1$ be two families of models, with ${\cal M}_0\subset {\cal M}_1$  and 
 $\dim({\cal M}_0)<\dim({\cal M}_1)$. 
 Consider the test of $H_0$ {\em within} $H_1$, opposing  $H_0$ : ``$f^M\in {\cal M}_0$" against 
 $H_1$ : ``$f^M\in {\cal M}_1$". 

\vspace{0.1in}

By construction, $K( f^D ||\hat{f}^{{\cal M}_0})\ge K(f^D || \hat{f}^{{\cal M}_1})$ since 
${\cal M}_1$ is a more general model than ${\cal M}_0$. Under $H_1$, their difference can be shown to follow asymptotically a chi-square distribution. Precisely, the 
{\em nested test} of $H_0$ within $H_1$ reads: ``{\em under the assumption that $H_1$ holds, rejects $H_0$ if}
\begin{eqnarray}
\label{nestedtest}
2n\: [K( f^D ||\hat{f}^{{\cal M}_0})-K( f^D ||\hat{f}^{{\cal M}_1})]\ge       
 \chi^2_{1-\alpha}\scriptstyle{[\dim({\cal M}_1)-\dim({\cal M}_0)]} \: \mbox{"}.
 \end{eqnarray}

\subsubsection{Example: quasi-symmetry, symmetry and marginal homogeneity}
\label{qqq191}
 Flows can be  represented by a square matrix $f_{jk}\ge0$ such that $\sum_{j=1}^m\sum_{k=1}^m f_{jk}=1$, with the representation ``$f_{jk}$~=~{\em proportion of units located at place $j$ at some time and at place $k$  some fixed time later}".

\vspace{0.1in}

A popular model for flows is the {\em quasi-symmetric class} $\mbox{\tt QS}$ (Caussinus 1966), known as the {\em Gravity model} in Geography (Bavaud 2002a) 
\begin{eqnarray*}
\mbox{\tt QS}=\{f\: |\: f_{jk}=\alpha_j\beta_k\gamma_{jk}\qquad\mbox{\small with }\: \gamma_{jk}=\gamma_{kj}\}
\end{eqnarray*}
where $\alpha_j$ quantifies the ``push effect",  $\beta_k$ the ``pull effect" and  $\gamma_{jk}$ the  ``distance deterrence function". 

\vspace{0.1in}

\noindent {\em Symmetric} and {\em marginally homogeneous} models constitute two popular alternative families, defined as 
\begin{eqnarray*}
\mbox{\tt S}=\{f\: |\: f_{jk}=f_{kj}\}
\qquad\qquad\qquad
\mbox{\tt MH}=\{f\: |\: f_{j\bullet}=f_{\bullet j}\}\:.
\end{eqnarray*}
Symmetric and quasi-symmetric ML estimates satisfy (see e.g. Bishop and al. (1975) or Bavaud (2002a))
\begin{eqnarray*}
\hat{f}^{\tt S}_{jk}=\frac12(f_{jk}^D+f_{kj}^D)\: \: 
\qquad 
\hat{f}^{\tt QS}_{jk}+\hat{f}^{\tt QS}_{kj}=f^D_{jk}+f^D_{kj}\: 
 \qquad
\hat{f}^{\tt QS}_{j\bullet}=f^D_{j\bullet} \qquad\: 
\hat{f}^{\tt QS}_{ \bullet k}=f^D_{ \bullet k}
\end{eqnarray*}
from which the values of $\hat{f}^{\tt QS}$ can be obtained iteratively. A similar yet more involved  procedure permits to obtain the 
marginal homogeneous estimates
$\hat{f}^{\tt MH}$.

\vspace{0.1in}

By construction, ${\tt S}\subset {\tt QS}$,  and the test (\ref{nestedtest})  consists in rejecting ${\tt S}$ (under the assumption that  {\tt QS} holds) if 
\begin{eqnarray}
\label{nestqsqsest}
2n\: [K( f^D ||\hat{f}^{\tt S})-K( f^D ||\hat{f}^{\tt QS})]\ge       
 \chi^2_{1-\alpha}\scriptstyle{[m-1]} \:. \end{eqnarray}
Noting that $\mbox{\tt S}=\mbox{\tt QS}\cap \mbox{\tt MH}$, 
 (\ref{nestqsqsest}) actually constitutes an alternative  testing procedure for {\tt QS}, avoiding the necessity of computing $\hat{f}^{\tt MH}$ (Caussinus 1996).

\subsubsection*{Example \ref{qqq191} continued: inter-regional migrations}
Relative entropies associated to Swiss inter-regional migrations flows 1985-1990 ($m=26$ cantons; see Bavaud (2002a)) are 
$K(f^D||\hat{f}^{\tt S})=.00115$ (with $\mbox{\small df}=325$) and   
$K(f^D||\hat{f}^{\tt QS})=.00044$ (with $\mbox{\small df}=300$). The difference is $.00071$
(with $\mbox{\small df}=25$ only)  and indicates that flows asymmetry is mainly produced by 
the  violation of marginal homogeneity (unbalanced flows)  rather than  the violation of quasi-symmetry. However, the sheer size of the sample $(n=6'039'313)$ leads,  at conventional significance levels,  to reject  all three models
$\mbox{\tt S}$ , $\mbox{\tt MH}$ and $\mbox{\tt QS}$.

 \subsection{Competition between simple hypotheses: Bayesian selection}
Consider the set of $q$ simple hypotheses ``$H_a$ : $f^M=g^a$ ", where $g^a\in S_m$ for $a=
1,\ldots, q$. In a Bayesian setting, denote by $P(H_a)=P(g^a)>0$  the prior probability of 
 hypothesis $H_a$, with 
$\sum_{a=1}^qP(H_a)=1$. The posterior probability $P(H_a|D)$ obtains from Bayes rule as
\begin{eqnarray*}
 P(H_a|D)=\frac{P(H_a)\: P(D|H_a)}{P(D)}\qquad\qquad\mbox{\small with}
\quad P(D)=\sum_{a=1}^qP(H_a) \: P(D|H_a)\:.
\end{eqnarray*}
Direct application of the asymptotic rate formula (\ref{typesform}) then yields
\begin{eqnarray}
\label{bsfbsf}
\mbox{\small $ P(g^a|f^D)\cong\frac{P(g^a)\: \exp(-n\: K(f^D||g^a))}{P(f^D)}$}
 \quad  \mbox{\small (Bayesian hypothesis selection formula)}
\end{eqnarray}
which shows, for $n\to \infty$, the posterior probability to be concentrated on the 
(supposedly unique) solution of 
 \begin{eqnarray*}
\hat{g}=\arg\min_{g^a}\: K(f^\ast||g^a)\qquad\qquad\mbox{\small where}
\quad f^\ast:=\lim_{n\to \infty}f^D\:.
\end{eqnarray*}
In other words, the asymptotically surviving model $g^a$   minimises  the 
 relative entropy $K(g^a||f^\ast)$ with respect to the 
 long-run empirical  distribution $f^\ast$, in accordance with the ML principle. 
 
 \vspace{0.1in}
 
For finite $n$, the relevant functional is $K(f^D||g^a))-\frac1n \ln P(g^a)$, where the  second term represents a prior penalty  attached to  hypothesis $H_a$. Attempts to generalize this framework to {\em families} of models ${\cal M}_a$ ($a=1,\ldots, q$) lie at the heart of the so-called {\em model selection procedures}, with the introduction of penalties (as in the  AIC, BIC, DIC, ICOMP, etc. approaches)  increasing with the number of free parameters $\mbox{\small dim}({\cal M}_a)$   (see e.g. Robert (2001)). In the alternative {\em minimum description length} (MDL) and
 {\em algorithmic complexity theory} approaches (see e.g. MacKay (2003) or  Li and Vitanyi (1997)), 
richer models necessitate a longer description and should be penalised accordingly. All those procedures, together with Vapnik's {\em Structural Risk Minimization} (SRM) principle (1995),  aim at controlling the problem of over-parametrization in statistical modelling.  We shall not pursue any further those matters, whose conceptual and methodological unification remains yet to accomplish. 

\subsubsection{Example: Dirichlet priors}
Consider the continuous {\em Dirichlet prior} $g\sim {\cal D}(\underline{\alpha})$, with  density  
$\rho(g|\underline{\alpha})=\frac{\Gamma(\alpha)}{\prod_j \Gamma(\alpha_j)}\prod_j g_j^{\alpha_j-1}$, normalised to unity  in $S_m$, where
$\underline{\alpha}=(\alpha_1,\ldots,\alpha_m)$ is a vector of   parameters with 
$\alpha_j>0$  and  $\alpha:=\sum_j \alpha_j$. Setting $\pi_j:=\alpha_j/\alpha=E(g_j|\underline{\alpha})$, Stirling approximation 
yields $ \rho(g|\underline{\alpha})\cong \exp(-\alpha K(\pi||g))$ for $\alpha$ large. 

Alfter observing the data
  $\underline{n}=(n_1,\ldots,n_m)$, the posterior distribution is well-known to be $ {\cal D}(\underline{\alpha}+\underline{n})$. Using $f_j^D=n_j/n$, one gets $ \rho(g|\underline{\alpha}+\underline{n})/\rho(g|\underline{\alpha})\cong \exp(-n K(f^D||g))$ for $n$ large,  as it must from   (\ref{bsfbsf}). Hence
\begin{eqnarray}
\label{kjsdsh}
\rho(g|\underline{\alpha}+\underline{n})\cong \exp[-\alpha  K(\pi||g)- n K(f^D||g)]\cong 
\exp[-(\alpha+n) K(f^{\mbox{\tiny post}}||g)]
\end{eqnarray}
 \begin{eqnarray}
\label{ksliuzsh}
\mbox{where}  \quad
f^{\mbox{\tiny post}}_j=E(g_j|\underline{\alpha}+\underline{n})=\lambda\:  \pi_j+(1-\lambda)f_j^D
 \quad  \mbox{\small with}\quad\lambda:=\frac{\alpha}{\alpha+n}\: .
\end{eqnarray}
(\ref{kjsdsh}) and  (\ref{ksliuzsh}) show  the parameter $\alpha$ to measure the strength of belief in the prior guess, measured in units of the sample size (Ferguson 1974).

\section{Maximum entropy}
\label{maoIAUS}
\subsection{Large deviations: Sanov's theorem}
 Suppose data to be {\em incompletely observed}, i.e. one only knows that $f^D\in {\cal D}$, where 
${\cal D}\subset S$ is a subset of the simplex $S$, the set of all possible distributions with $m$ modalites. Then, for an i.i.d. process, a theorem due to Sanov (1957) says that,   for sufficiently regular ${\cal D}$, the asymptotic rate of the probability that $f^D\in {\cal D}$ under model $f^M$ decreases exponentially as 
 \begin{eqnarray}
 \label{maxent00}
 P(f^D\in {\cal D}|f^M) \cong  \exp(-n\: K(\tilde{f}^{\cal D}||f^M))\:
 \mbox{\small where}\: 
\tilde{f}^{\cal D} := \arg\min_{f\in {\cal D}} K(f||f^M)\:.
\end{eqnarray}
$\tilde{f}^{\cal D}$ is the so-called 
 {\em maximum entropy} (ME) solution, that is the most probable empirical distribution under the prior model  $f^M$ and the knowledge that  $f^D\in {\cal D}$. Of course,  $\tilde{f}^{\cal D}=f^M$ if $f^M\in  {\cal D}$.

 \subsection{On the nature of the maximum entropy solution}
 \label{natureme}
 When the prior is uniform ($f_j^M=1/m$), then 
\begin{eqnarray*}
 K(f^D||f^M)=\ln m - H(f^D)
 \end{eqnarray*}
and {\em minimising} (over $f\in {\cal D}$) the relative entropy $K(f||f^M)$  amounts in 
{\em maximising} the entropy $H(f^D)$ (over $f\in {\cal D}$).
   \vspace{0.1in}

For decades (ca. 1950-1990), the ``maximum entropy" principle, also called ``minimum  discrimination information (MDI) principle" by Kullback (1959), has largely been used in science and engineering as a  {\em first-principle}, ``maximally non-informative" method of generating {\em models}, maximising our ignorance (as represented by the entropy) under our available knowledge ($f\in {\cal D}$) (see in particular Jaynes (1957), (1978)). 

\vspace{0.1in}

However, (\ref{maxent00}) shows the maximum entropy construction to be justified from Sanov's theorem, and to result form the minimisation of the {\em first} argument of the relative entropy, which points towards the {\em empirical} (rather than theoretical) nature of the latter. In the present setting, $\tilde{f}^{\cal D}$ appears as  the most likely data reconstruction  under  the prior model  and the incomplete observations (see also section \ref{pythagor}). 

 \subsubsection{Example: unobserved category}
\label{qqq555}
Let $f^M$ be given and suppose one knows that a category, say  $j=1$, has {\em not} occured. Then
\begin{eqnarray*}
  \tilde{f}^{\cal D}_j=\left\{\begin{array}{cl}0 & \mbox{for  $j= 1$}\\  \frac{f_j^M}{1-f^M_1} & \mbox{for  $j> 1$}
\end{array}\right.\qquad\qquad \mbox{\small and}\quad  K(\tilde{f}^{\cal D}||f^M)=-\ln (1-f_1^M),  
\end{eqnarray*}
whose finiteness (for $f^M_1<1$) contrasts the behavior  $K(f^M|| \tilde{f}^{\cal D})=\infty$ (for $f^M_1>0$). See example \ref{qqq34} f).

 \subsubsection{Example: coarse grained observations}
\label{qqq556}
Let $f^M$ be a given distribution with categories $j=1,\ldots,m$. Let $J=1,\ldots,M<m$ denote {\em groups of categories}, and suppose that observations are aggregated or coarse-grained, i.e. of the form
\begin{eqnarray*}
{\cal D}=\{\: f^D\:  |\: \sum_{j\in J}f_j^D\stackrel{!}{=}F_J^D\quad J=1,\ldots,M\: \}\:.
\end{eqnarray*}
Let $J(j)$ denote the group to which $j$ belongs. The ME distribution then reads  (see (\ref{cocococ}) and example \ref{qqq087})
\begin{eqnarray}
\label{cogrob}
\tilde{f}^{\cal D}_j=f_j^M   
  \: \frac{F_{J(j)}^D}{F_{J(j)}^M}\: \: \:  \mbox{\small where}\: \: \: F_J^M:=\sum_{j\in J}f_j^M
  \: \: \: \mbox{\small and}\:  \: \: K(\tilde{f}^{\cal D}||f^M)=K(F^D||F^M) .
\end{eqnarray}

\subsubsection{Example: symmetrical observations}
\label{qqq557}
 Let $f_{jk}^M$ be a given joint model for {\em square} distributions ($j,k=1,\ldots,m$). Suppose one knows the data distribution to be symmetrical, i.e. 
\begin{eqnarray*}
{\cal D}=\{f\: |\: f_{jk}^D=f_{kj}^D\: \} \:.
\end{eqnarray*}

Then
\begin{eqnarray*}
  \tilde{f}^{\cal D}_{jk}=\frac{\sqrt{f_{jk}^M\: f_{kj}^M}}{Z}\qquad\qquad\mbox{\small where}\qquad
Z:=\sum_{jk}\sqrt{f_{jk}^M\: f_{kj}^M}
\end{eqnarray*}
which is contrasted with the result $\hat{f}^{\cal M}_{jk}=\frac12(f_{jk}^D+f_{kj}^D)$ of example 
\ref{qqq191} (see section \ref{lefd}).

\subsection{``Standard" maximum entropy: linear constraint}
\label{stlico}
\begin{figure}
 \begin{center}
\includegraphics[width=2.5in]{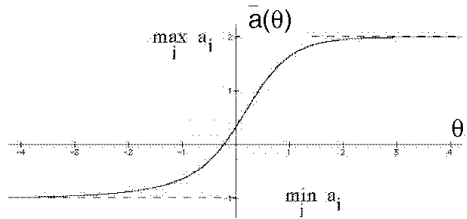}
\caption{typical behaviour of $\bar{a}(\theta)$}
\label{stamxe}
\end{center}
\end{figure}

Let ${\cal D}$ be determined by a {\em linear constraint} of the form
\begin{eqnarray*}
{\cal D}=\{ f\: |\: \sum_{j=1}^m f_j a_j= \bar{a}\: \}\qquad\qquad \mbox{\small with}\qquad  \min_ja_j\le \bar{a}\le \max_j a_j
\end{eqnarray*}
that is, one knows the empirical average  of some quantity $\{a_j\}_{j=1}^m$ to be fixed to $\bar{a}$. Minimizing 
over $f\in S$ the functional 
\begin{eqnarray}
\label{oiu}
K(f||f^M)+\theta A(f)\qquad\qquad\qquad  A(f):=\sum_{j=1}^m f_j a_j
\end{eqnarray}
\begin{eqnarray}
\label{maxenpri}
 \mbox{yields}\qquad\qquad\tilde{f}^{\cal D}_{j}=\frac{f_j^M\: \exp(\theta a_j)}{Z(\theta)}
  \qquad\qquad
  Z(\theta):=\sum_{k=1}^m f_k^M\: \exp(\theta a_k)
  \end{eqnarray}
where the Lagrange multiplier $\theta$ is determined by the constraint $\bar{a}(\theta):=\sum_j  \tilde{f}^{\cal D}_{j}(\theta)\: a_j \stackrel{!}{=}\bar{a}$ (see figure \ref{stamxe}).

\subsubsection{Example: average value of a dice}
\label{qqq333}
Suppose one believes a dice to be fair ($f_j^M=1/6$),  and one is told that the empirical average of its face values is say $\bar{a}=\sum_j f_j^D\: j =4$, instead of $\bar{a}=3.5$ as expected. The value of $\theta$ in (\ref{maxenpri})  insuring $\sum_j \tilde{f}^{\cal D}_{j}\: j=4$ turns out to be 
 $\theta=0.175$, insuring $\sum_j \tilde{f}^{\cal D}_{j}\: j=4$, as well as  $\tilde{f}^{\cal D}_{1}=0.10$, $\tilde{f}^{\cal D}_{2}=0.12$, 
$\tilde{f}^{\cal D}_{3}=0.15$, 
$\tilde{f}^{\cal D}_{4}=0.17$, 
$\tilde{f}^{\cal D}_{5}=0.25$, 
$\tilde{f}^{\cal D}_{6}=0.30$
(Cover and Thomas  (1991) p. 295).

\subsubsection{Example: Statistical Mechanics}
\label{qqq444}
An interacting particle system can occupy $m>>1$ configurations $j=1,\ldots, m$, a priori equiprobable ($f^M_j=1/m$), with corresponding energy $E_j$. 
Knowing the average energy to be $\bar{E}$, the resulting ME solution (with $\beta:=-\theta$) is the {\em Boltzmann-Gibbs distribution}
\begin{eqnarray}
\label{bolgib}
  \tilde{f}^{\cal D}_{j}=\frac{\exp(- \beta  E_j)}{Z(\beta)}
  \qquad\qquad 
  Z(\beta):=\sum_{k=1}^m   \exp(-\beta E_k)
\end{eqnarray}
minimising the {\em free energy} $F(f):=E(f)-TH(f)$, obtained (up to a constant term)  by multiplying the functional (\ref{oiu}) by the {\em temperature} $T:=1/\beta=-1/\theta$. Temperature plays the role of an arbiter determining the trade-off between the contradictory objectives of energy minimisation and entropy maximisation: 
\begin{enumerate}
  \item[$\bullet$]  at  high temperatures   $T\to\infty$ (i.e. $\beta\to 0^+$), the Boltzmann-Gibbs distribution
   $ \tilde{f}^{\cal D}$ becomes uniform and the entropy $H( \tilde{f}^{\cal D})$ maximum
   (fluid-like organisation of the matter).
    \item[$\bullet$]  at  low  temperatures   $T\to 0^+$ (i.e. $\beta\to \infty$), the Boltzmann-Gibbs distribution
     $ \tilde{f}^{\cal D}$ becomes concentrated on the {\em ground states} $j_-:=\arg\min_j E_j$, making 
 the average energy  $E( \tilde{f}^{\cal D})$ minimum (crystal-like organisation of the matter). 
\end{enumerate}

  \subsubsection*{Example \ref{qqq191}, continued: quasi-symmetry}
  ME approach to gravity modelling consists in considering flows constrained by $q$ linear constraints of the form
\begin{eqnarray*}
{\cal D}=\{ f\: |\: \sum_{j,k=1}^m f_{jk} a_{jk}^\alpha= \bar{a}^\alpha\: \qquad\alpha=1,\ldots,q \} \end{eqnarray*}
such that, typically
\begin{itemize}
 \item[1)] $a_{jk}:=d_{jk}=d_{kj}$ (fixed average trip distance, cost or time $d_{jk}$) 
 \item[2)] $a^\alpha_{jk}:=\delta_{j\alpha}$  (fixed origin profiles, $\alpha=1,\ldots,m$ )
\item[3)] $a^\alpha_{jk}:=\delta_{\alpha k}$  (fixed destination profiles, $\alpha=1,\ldots,m$)
\item[4)] $a_{jk}:=\delta_{jk}$ (fixed proportion of   stayers)
\item[5)] $a_{jk}:=\delta_{j\alpha}-\delta_{\alpha k}$ (balanced flows, $\alpha=1,\ldots,m$)
\end{itemize}
 Constraints 1) to 5) (and linear combinations of them) yield all the ``classical  Gravity models" proposed in Geography,  such as the {\em exponential decay model} (with $f^M_{jk}=a_j\: b_k$): 
\begin{eqnarray*}
\tilde{f}_{jk}^{\cal D}=\alpha_j\beta_k\exp(-\beta d_{jk})
\end{eqnarray*}
Moreover, if  the prior $f^M$ is quasi-symmetric, so is  $\tilde{f}^{\cal D}$ under the above constraints (Bavaud 2002a). 
 
\section{Additive decompositions}
\subsection{Convex and exponential families of distributions}
\label{lefd}

 {\bf  Definition:} a family ${\cal F}\subset S$ of distributions is a  {\bf convex} family  iff 
  \begin{eqnarray*}
f,g\in {\cal F}\Rightarrow \lambda f +(1-\lambda) g\in {\cal F}\qquad\qquad \forall \: \lambda\in [0,1]
\end{eqnarray*}
Observations typically involve the identification  of  merged categories, and the corresponding 
empirical distributions are coarse grained, that is determined through aggregated values $F_J:=\sum_{j\in J} f_j$ only. Such coarse grained distributions 
form a convex family (see table \ref{tablece}). More generally, linearly constrained distributions (section \ref{stlico}) are convex.  Distributions (\ref{npoptrr}) belonging to the optimal Neyman-Pearson  regions $W$ (or $W^c$), posterior  distributions (\ref{ksliuzsh}) as well as marginally homogeneous distributions (example \ref{qqq191}) provide other examples of convex  families. 

\begin{table}
  \centering
  \begin{tabular}{|ccccc|}
\hline
Family ${\cal F}$ & characterization & remark & {\small convex} & {\small expon.} \\  \hline\hline 
deficient & $f_1=0$ & & {\small{yes}} & {\small{yes}} \\ \hline
deterministic & $f_1=1$ & & {\small{yes}} & {\small{yes}} \\ \hline
coarse grained & $\sum_{j\in J}f_j=F_J$ & & {\small{yes}} & {\small{no}} \\ \hline
mixture & $f_{j}=f_{(Jq)}=\rho_q h_{J}^q$     & $\{h_{J}^q\}$ fixed\hspace{0.8cm}  & {\small{yes}} & {\small{yes}} \\
\hline
mixture & $f_{j}=f_{(Jq)}=\rho_q h_{J}^q$     & $\{h_{J}^q\}$ {\small adjustable} & {\small{no}} & {\small{yes}} \\
\hline
independent & $f_{jk}=a_j b_k$ & & {\small{no}} & {\small{yes}} \\
 \hline 
{\small marginally homog.} & $f_{j\bullet}=f_{\bullet j}$ & square tables & {\small{yes}} & {\small{no}} \\ \hline
symmetric & $f_{jk}=f_{kj}$ & square tables & {\small{yes}} & {\small{yes}} \\ \hline
quasi-symmetric & {\small $f_{jk}=a_j b_k c_{jk}$, $c_{jk}=c_{kj}$} & square tables & {\small{no}} & {\small{yes}} \\ \hline
\end{tabular}
 \caption{some convex  and/or  exponential  families}\label{tablece}
\end{table}

\vspace{0.1in}

\noindent {\bf  Definition:} a family ${\cal F}\subset S$ of distributions is an {\bf exponential} family iff 
  \begin{eqnarray*}
f,g\in {\cal F}\Rightarrow\frac{f^\mu g^{1-\mu}}{Z(\mu)}\in {\cal F}\qquad\mbox{\small where}\quad Z(\mu):=\sum_{j=1}^m f_j^\mu g_j^{1-\mu}\qquad\forall \: \mu\in [0,1]
\end{eqnarray*}
 Exponential families are a favorite object of classical statistics. Most  classical discrete or continuous probabilistic models (log-linear, multinomial, Poisson, Dirichlet, Normal, Gamma, etc.) constitute exponential families. Amari (1985) has developed a local parametric  characterisation of exponential and convex families in a differential geometric framework. 

\subsection{Factor analyses}
Independence models are exponential but not convex (see table \ref{tablece}): the weighted sum of independent distributions is not independent in general. Conversely, non-independent distributions can be decomposed as a sum of (latent) independent terms through {\em factor analysis}. 
The spectral decomposition of the chi-square producing the {\em factorial correspondence analysis} of contingency tables turns out to be exactly applicable on 
mutual information (\ref{indeptrio}) as well,  yielding an ``entropic" alternative to (categorical) factor analysis (Bavaud 2002b). 

  \vspace{0.1in}
  
  {\em Independent component analysis} (ICA) aims at determining the linear transformation of  multivariate (continuous) data making them as independent as possible. In contrast to {\em principal component analysis}, limited to the second-order statistics associated to gaussian models, ICA attempts to  take into account higher-order dependencies occurring in the mutual information between variables, and extensively relies  on information-theoretic  principles, as developed in Lee et al. (2000) or Cardoso (2003) and references therein.

 \subsection{Pythagorean theorems}
 \label{pythagor}
 The following results, sometimes referred to as the {\em Pythagorean theorems of IT},  provide an exact additive decomposition of the relative entropy: 
 
 \vspace{0.1in}

{\bf  Decomposition theorem for convex families:}  if ${\cal D}$ is a convex family, then 
\begin{eqnarray}
\label{addD}
K(f||f^M)= K(f||\tilde{f}^{\cal D})+K(\tilde{f}^{\cal D}||f^M) \qquad\mbox{for any}\:  f\in {\cal D}
\end{eqnarray}
where $\tilde{f}^{\cal D}$ is the ME distribution for ${\cal D}$ with prior $f^M$.

\vspace{0.1in}

{\bf  Decomposition theorem for exponential families:}  if  ${\cal M}$ is an exponential family, then 
\begin{eqnarray}
\label{addM}
K(f^D||g)= K(f^D||\hat{f}^{\cal M})+K(\hat{f}^{\cal M}|| g) \qquad\mbox{for any}\:  g\in {\cal M}
\end{eqnarray}
where $\hat{f}^{\cal M}$ is the ML distribution for ${\cal M}$ with data $f^D$.

\vspace{0.1in}

{\bf Sketch of the proof of (\ref{addD})} (see e.g. Simon 1973): if ${\cal D}$ is convex with $\mbox{\small dim}( {\cal D})=\mbox{\small dim}( {\cal S})-q$, its elements are of the form
${\cal D}=\{f\: |\: \sum_j f_j a_j^\alpha=a_0^\alpha\: \mbox{\small for}\:  \alpha=1,\ldots,q\}$, which implies the maximum entropy solution to be of the form $\tilde{f}^{\cal D}_j=\exp(\sum_\alpha\lambda_\alpha a_j^\alpha) f_j^M/Z(\lambda)$. 
Substituting this expression and using  $\sum_j f_j a_j^\alpha=\sum_j \tilde{f}^{\cal D}_j a_j^\alpha$ proves (\ref{addD}).

\vspace{0.1in}

{\bf Sketch of the proof of (\ref{addM})} (see e.g. Simon 1973): if  ${\cal M}$ is exponential with $\mbox{\small dim}( {\cal M})=r$, its elements are of the
 form $f_j=\rho_j\exp(\sum_{\alpha=1}^r\lambda_\alpha a_j^\alpha) /Z(\lambda)$ (where the {\em 
 partition function} $Z(\lambda)$ insures the normalisation),  containing $r$ free non-redundant  parameters $\lambda\in \mathbb{R}^r$. Substituting this expression and using the optimality condition
 $\sum_j\hat{f}^{\cal M}_ja_j^\alpha=\sum_j f^D_ja_j^\alpha$ for all $\alpha=1,\ldots,r$ 
proves (\ref{addM}).

 \vspace{0.1in}
 
Equations (\ref{addD}) and (\ref{addM}) show that 
   $\tilde{f}^{\cal D}$ and $\hat{f}^{\cal M}$ can both occur as {\em left} and {\em right} arguments of the relative entropy, underlining their somehow {\em hybrid} nature, {\em intermediate between data and models} (see section \ref{natureme}).

\subsubsection{Example: nested tests}
Consider two exponential families ${\cal M}$ and  ${\cal N}$ with 
${\cal M}\subset {\cal N}$. Twofold application  of 
   (\ref{addM}) demonstrates the identity
\begin{eqnarray*}
 K( f^D ||\hat{f}^{{\cal M}})-K( f^D ||\hat{f}^{{\cal N}})=K(\hat{f}^{\cal N}|| \hat{f}^{\cal M}) 
 \end{eqnarray*} 
occuring in nested tests such as (\ref{nestqsqsest}). 
    
\subsubsection{Example: conditional independence in three-dimensional tables}
\label{qqq129}
\label{ciitdt}
Let $f_{ijk}^D:=n_{ijk}/n$ with $n:=n_{\bullet\bullet\bullet}$ be the empirical distribution associated to the 
 $n_{ijk}$ = ``{\em number of individuals in the category $i$ of $X$, $j$ of $Y$ and $k$ of $Z$} ".
 Consider the families of models
 \begin{eqnarray*}
 {\cal L} \:  = &\{ f\in S\: |\: f_{ijk}=a_{ij}b_k \} & =  \:   \{ f\in S\: |\: \ln f_{ijk}=\lambda +\alpha_{ij}+\beta_k \} \\
{\cal M} \:   = &\{ f\in S\: |\: f_{\bullet jk}=c_{i}d_k \}& =    \: \{ f\in S\: |\: \ln f_{\bullet jk}=\mu +\gamma_j +\delta_k
 \} \\
{\cal N}  \:  = &\{ f\in S\: |\: f_{ijk}=e_{ij}h_{jk} \}& = \:    \{ f\in S\: |\: \ln f_{ijk}=\nu +\epsilon_{ij}+\eta_{jk} \}\: .
\end{eqnarray*}
Model ${\cal L}$ expresses that $Z$ is independent from $X$ and $Y$ (denoted $Z\perp (X,Y)$).  
Model ${\cal M}$ expresses that $Z$ and $Y$ are independent  ($Y\perp Z$). Model
${\cal N}$ expresses that, conditionally to $Y$,  $X$ and $Z$ are  independent  ($X\perp Z|Y$). 
Models ${\cal L}$ and ${\cal N}$ are exponential (in $S$), and ${\cal M}$ is exponential in the space of joint distributions on $(Y,Z)$. They constitute well-known examples  of {\em log-linear models}  (see e.g. Christensen (1990)). 

\vspace{0.1in}

Maximum likelihood estimates and  associated relative entropies obtain as (see example \ref{qqq088})
\begin{eqnarray*}
\hat{f}^{\cal L}_{ijk}=f^D_{ij\bullet}f^D_{\bullet \bullet  k}
\: & \Rightarrow &\: 
K(f^D||\hat{f}^{\cal L})= H_D(XY)+H_D(Z)-H_D(XYZ)
\\
 \hat{f}^{\cal M}_{ijk}=
\frac{f_{ijk}^D}{f_{\bullet jk}^D}\: f^D_{\bullet j\bullet}\:  f^D_{ \bullet\bullet k}
\:  & \Rightarrow & \: 
K(f^D||\hat{f}^{\cal M})= H_D(Y)+H_D(Z)-H_D(YZ)
 \\
 \hat{f}^{\cal N}_{ijk}=\frac{f^D_{ij\bullet}\: 
f^D_{ \bullet jk}}{f^D_{ \bullet j\bullet }}
\:  & \Rightarrow & \: 
\mbox{\small $K(f^D||\hat{f}^{\cal N})= H_D(XY)+H_D(YZ)-H_D(XYZ)-H_D(Y)$}
\end{eqnarray*}
and permit to test the corresponding models as in (\ref{testcomp}). As a matter of fact, the present  example  illustrates another aspect of exact 
decomposition, namely ${\cal L}={\cal M}\cap {\cal N}$
\begin{eqnarray*}
f_{ijk}^D\hat{f}^{\cal L}_{ijk}=\hat{f}^{\cal M}_{ijk} \hat{f}^{\cal N}_{ijk}\quad
K(f^D||\hat{f}^{\cal L})=K(f^D||\hat{f}^{\cal M})+K(f^D||\hat{f}^{\cal N})\quad
\scriptstyle{\mbox{df}}^{\cal L}=\scriptstyle{\mbox{df}}^{\cal M}+\scriptstyle{\mbox{df}}^{\cal N}
\end{eqnarray*}
where $\scriptstyle{\mbox{df}}$ denotes the appropriate degrees of freedom for the chi-square test~(\ref{testcomp}).

 \subsection{Alternating minimisation and the EM algorithm}
  \subsubsection{Alternating minimisation} 
 Maximum likelihood and maximum entropy are particular cases of the general problem
 \begin{eqnarray}\label{genprob}
 \min_{f\in {\cal F}} \min_{g\in {\cal G}}K(f||g)\: .
\end{eqnarray}
 {\em Alternating minimisation} consists in defining recursively 
 \begin{eqnarray}
f^{(n)} & := & \arg\min_{f\in {\cal F}} K(f||g^{(n)}) \label{estep}\\
  g^{(n+1)}  & := & \arg\min_{g\in {\cal G}} K(f^{(n)}||g) \:. \label{mstep}
\end{eqnarray}
Starting  with some $g^{(0)}\in{\cal G}$ (or 
some $f^{(0)}\in{\cal F}$), and for  ${\cal F}$ and ${\cal G}$ convex,  $K(f^{(n)}||g^{(n)})$ converges towards (\ref{genprob})
 (Csisz{\'a}r (1975); Csisz{\'a}r and Tusn{\'a}dy, 1984). 

 \subsubsection{The EM algorithm}
Problem (\ref{estep}) is easy to solve
 when   ${\cal F}$ is the coarse grained family $\{f \: |\: \sum_{j\in J}f_j=F_J\}$, with solution   (\ref{cogrob})
$f^{(n)}_j =g^{(n)}_j\: F_{J(j)}/ G^{(n)}_{J(j)}$  
and  the result 
$K(f^{(n)}||g^{(n)})=K(F||G^{(n)})$ (see example \ref{qqq556}). 

The present situation describes  {\em incompletely observed data}, in which $F$ only (and not $f$) is known, 
with corresponding model 
$G(g)$ in ${\cal M}:=\{G\: |\: G_J=\sum_{j\in  J} g_j\: \mbox{\small and}\: g\in {\cal G}\}$. Also
\begin{eqnarray*}
\min_{G\in {\cal M}}K(F||G)&=&\min_{g\in {\cal G}}K(F||G(g))=\min_{g\in {\cal G}}\min_{f\in {\cal F}}K(f||g)\\
&=&
\lim_{n\to\infty}K(f^{(n)}||g^{(n)})=\lim_{n\to\infty} K(F||G^{(n)})
\end{eqnarray*}
which shows $G^{(\infty)}$ to be the solution  of $\min_{G\in {\cal M}}K(F||G)$. This particular version of the alternating minimisation procedure is known as 
the {\em EM algorithm} in the literature (Dempster et al. 1977), where (\ref{estep}) is referred to as the ``{\em  expectation step}" and (\ref{mstep}) as the ``{\em  maximisation step}".

\vspace{0.1in}

Of course, the above procedure is fully operational provided  (\ref{mstep})  can also be easily solved. This occurs for instance for {\em finite-mixture models} determined by $c$ {\em fixed} distributions $h_J^q$ (with 
$\sum_{J=1}^mh_J^q=1$ for $q=1,\ldots,c$), such that the categories  
 $j=1,\ldots, m$ read as 
 product categories of the form $j=(J,q)$  with 
\begin{eqnarray*}
g_{j}=g_{(Jq)}=\rho_q\: h_{J}^q\qquad\quad \rho_q\ge0\qquad\quad
\sum_{q=1}^c \rho_q=1 
\qquad\quad 
G_J=\sum_q \rho_q h_{J}^q\end{eqnarray*}
where the ``mixing proportions" $\rho_q$ are freely adjustable. Solving (\ref{mstep}) yields
\begin{eqnarray*}
\rho_q^{(n+1)}=\sum_J f_{(Jq)}^{(n)}=
\rho_q^{(n)}\: \sum_J\frac{h_{J}^q\: F_J}{\sum_r h_{J}^r\: \rho_r^{(n)}} 
\end{eqnarray*}
which converges towards  the optimal  mixing proportions $\rho_q^{(\infty)}$, unique since ${\cal G}$ is convex. Continuous versions of the algorithm (in which $J$ represents a position in an Euclidean space)  
generate the so-called {\em soft clustering} algorithms, which can be further restricted to the hard clustering and  $K$-means algorithms. However, the distributions $h_J^q$ used in the latter cases generally contain additional {\em adjustable} parameters (typically the mean and the covariance matrix of normal distributions), which  break  down the convexity of ${\cal G}$ and cause the algorithm to converge towards {\em local minima}.

 \section{Beyond independence: Markov chain models and texts}
 \label{qqq1}
As already proposed by Shannon (1948), the independence formalism can be extended to stationary dependent sequences, that is on 
categorical time series or  ``textual" data $D=x_1x_2\ldots x_n$, such as

\vspace{0.1in}

$D$={\tt 
bbaabbaabbbaabbbaabbbaabbaabaabbaabbaabbaabbaabbaabbaabb

aabaabbaabbbaabaabaabbbaabbbaabbaabbaabbaabaabbbaabbbaabbaa

baabaabbaabaabbaabbaabbbaabbaabaabaabbaabbbbaabbaabaabaabaa

baabaabaabbaabbaabbaabbbbaab}$\quad.$

\vspace{0.1in}

In this context, each occurence $x_i$ constitutes a {\em letter} taking values $\omega_j$ in a state space $\Omega$, the {\em alphabet}, of cardinality $m=|\Omega|$.    A sequence of $r$ letters $\alpha:=\omega_1\ldots \omega_r\in \Omega^r$ is an {\em $r$-gram}.  In our example, 
$n=202$, $\Omega=\{\mbox{\tt a}, \mbox{\tt b}\}$, $m=2$, $\Omega^2=\{\mbox{\tt aa},\mbox{\tt ab}, \mbox{\tt ba},\mbox{\tt bb}\}$, etc.  
  
  \subsection{Markov chain models}
A {\em Markov chain model of order $r$} is specified by the {\em conditional} probabilities
\begin{eqnarray*}
f^M(\omega|\alpha)\ge0\qquad  \omega\in \Omega\qquad\alpha\in \Omega^r\qquad
\sum_{\omega\in \Omega}f^M(\omega|\alpha)=1\:.
\end{eqnarray*}
$f^M(\omega|\alpha)$ is the probability that the symbol following the $r$-gram $\alpha$ is $\omega$.
It obtains from the {\em stationary distributions} $f^M(\alpha\omega)$ and $f^M(\alpha)$ as
 \begin{eqnarray*}
 f^M(\omega|\alpha)=\frac{f^M(\alpha\omega)}{f^M(\alpha)}\:
\end{eqnarray*}

\vspace{0.1in}

The  set ${\cal M}_r$ of  models of order $r$ constitutes an exponential family, nested as  
$ {\cal M}_r\subset {\cal M}_{r+1}$ for all $r\ge0$. In particular, 
${\cal M}_0$ denotes the independence models, and 
${\cal M}_1$ the ordinary (first-order) Markov chains. 

\vspace{0.1in}

The corresponding empirical distributions $f^D(\alpha)$ give the relative proportion of $r$-grams
$\alpha\in \Omega^r $ in the text $D$. They obtain as 
\begin{eqnarray*}
f^D(\alpha):=\frac{n(\alpha)}{n-r+1}
\qquad\qquad\mbox{\small with}\qquad\sum_{\alpha\in \Omega^r}f^D(\alpha)=1
\end{eqnarray*}
where $n(\alpha)$ counts the number of occurrences of $\alpha$ in $D$. In  the above example, the tetragrams counts are  for instance:
 
\begin{center}
\begin{tabular}{|c|c||c|c||c|c|}\hline
$\alpha$ & $n(\alpha)$  & $\alpha$ & $n(\alpha)$ & $\alpha$ & $n(\alpha)$ \\  \hline\hline
aaaa &  0  &
aaab & 0 &
aaba & 16   \\ \hline
aabb & 35   &
abaa & 16    &
abab & 0   \\ \hline
abba & 22      &
abbb & 11    & baaa & 0 \\ \hline
baab & 51 &
baba & 0     &
babb & 0   \\ \hline
bbaa & 35   &
 bbab & 0    &
bbba & 11     \\ \hline
bbbb & 2     & & &
{\bf total} & {\bf 199}    \\ \hline
\end{tabular}
\end{center}

 \subsection{Simulating a sequence}
 \label{qqq2}
 Under the assumption  that a text follows  a $r$-order model ${\cal M}_r$,  empirical distributions $f^D(\alpha)$  (with $\alpha\in \Omega^{r+1}$) converge for $n$ large to  $f^M(\alpha)$. The latter define in turn $r$-order transition probabilities, allowing the generation of new texts, started from the stationary distribution.

 \subsubsection{Example}
The following sequences are generated form the empirical probability transitions of the 
{\em Universal declaration
of Human Rights},  of length $n=8'149$ with  $m=27$ states (the alphabet + the blank, without punctuation): 

\vspace{0.1in}

{\bf $r=0$} (independent process)

{\tt  iahthire edr  pynuecu d lae mrfa  ssooueoilhnid nritshfssmo 

nise 
yye noa it eosc
e lrc jdnca tyopaooieoegasrors c hel 

niooaahettnoos rnei s  sosgnolaotd t atiet }

\vspace{0.1in}

{\bf $r=1$} (first-order Markov chain) 

{\tt erionjuminek in l ar hat arequbjus st d ase scin ero tubied 

pmed beetl
equly shitoomandorio tathic wimof tal ats evash 

indimspre tel sone aw 
onere pene e ed uaconcol mo atimered }

\pagebreak 
{\bf $r=2$} (second-order Markov chain) 
 
 {\tt mingthe rint son of the frentery and com andepent the halons 
 
 hal 
to coupon efornitity the rit noratinsubject will the the 

in priente  hareeducaresull ch infor aself and evell}
 
 \vspace{0.1in}

{\bf $r=3$} (third-order Markov chain) 

{\tt law socience of social as the right or everyone held

 genuinely available sament of his no one may be
enties the 

right in the cons as the 
right to equal co one soveryone}

\vspace{0.1in}

{\bf $r=4$} (fourth-order Markov chain)

{\tt are endowed with other means of full equality and to law no one

 is the right to choose of the detent
to arbitrarily in science 

with pay for through freely choice work}

 \vspace{0.1in}

{\bf $r=9$} (ninth-order Markov chain)

{\tt democratic society and is entitled without interference

 and to seek receive and impartial tribunals
for acts violating

 the fundamental rights
indispensable for his}

 \vspace{0.1in}
 
 Of course,  empirical distributions are expected to accurately estimate model distributions for 
  $n$ large enough, or equivalently for $r$ small enough, typically for 
\begin{eqnarray*}
r< r_{\max}:= \frac12\frac{\ln n}{\ln m}. 
\end{eqnarray*}
Simulations with $r$ 
above about $r_{\max}$ (here roughly equal to 2) are over-parameterized:  the number of parameters to be estimated exceeds the sample abilities to do so, and simulations  replicate fragments of the initial text rather than typical $r$-grams occurences of  written English in general, providing a vivid illustration of the  
{\em  curse of dimensionality} phenomenon.

\subsection{Entropies  and entropy rate}
The {\em $r$-gram entropy} and  the {\em conditional entropy of order $r$} associated to a (model or empirical) distribution $f$ are defined by 
\begin{eqnarray*}
H_r(f): = -\sum_{\alpha\in \Omega^r}f(\alpha)\: \ln f(\alpha) =  H(X_1,\ldots, X_r)\end{eqnarray*}
\begin{eqnarray*}
\mbox{\small $\displaystyle h_{r+1}(f): = -\sum_{\alpha\in \Omega^r} f(\alpha)  \sum_{\omega\in \Omega}f(\omega|\alpha)\: \ln f(\omega|\alpha)
 = H_{r+1}(f)-H_r(f)= H(X_{r+1}|X_1,\ldots, X_{r})\ge0 \: .$}
\end{eqnarray*}
The quantity $h_{r}(f)$ is non-increasing in $r$. Its limit defines the {\em entropy rate},  measuring 
the conditional uncertainty on the next symbol  knowing  the totality of past occurrences:
 \begin{eqnarray*}
h(f):=\lim_{r\to \infty} h_r(f)=\lim_{r\to \infty} \frac{H_r(f)}{r}\qquad\qquad\mbox{entropy rate.}
\end{eqnarray*}
By construction, $0\le h(f)\le \ln m$, and the so-called {\em redundancy} $R:=1-({h}/{\ln m})$ satisfies $0\le R\le1$.

The entropy rate measures the randomness of the stationary process:  $h(f)=\ln m$ (i.e. $R=1$) characterizes a maximally random process is,  that is a  dice model with uniform distribution. The process is ultimately deterministic iff $h(f)=0$ (i.e. $R=0$).

Shannon's estimate of the entropy rate of the written English on  $m=27$ symbols is about $h=1.3$ bits per letter, that is $h=1.3\times\ln 2= 0.90$ nat, corresponding to 
 $R=0.73$: hundred pages of written English are in theory compressible without loss to $100-73=27$ pages. Equivalently, using an alphabet containing $\exp(0.90)= 2.46 $ symbols only (and 
 the same number of pages) 
  is in principle sufficient to code the text without loss.

\subsubsection{Example: entropy rates for ordinary Markov chains} 
For a regular Markov chain of order 1 with transition matrix $W=(w_{jk})$ and stationary distribution $\pi_j$, one gets 
\begin{eqnarray*}
h_1=-\sum_j\pi_j\ln  \pi_j\ge h_2=h_3=\dots=-\sum_j\pi_j \sum_kw_{jk}\ln w_{jk}=h\: .
\end{eqnarray*}
Identity  $h_1=h$  holds iff $w_{jk}=\pi_k$, that is if the process is of order $r=0$. Also, $h\to 0$ iff 
$W$ tends to a permutation, that is iff the process becomes deterministic. 

\subsection{The asymptotic rate  for Markov chains}
Under the assumption of a model $f^M$ of order $r$, the probability to observe $D$ is   
\begin{eqnarray*}
\mbox{\small $\displaystyle P(D|f^M)\cong \prod_{i=1}^n P(x_{i+r}|x_i^{i+r-1})\cong\prod_{\omega\in\Omega}\prod_{\alpha\in\Omega^r}
f^M(\omega|\alpha)^{n(\alpha\omega)}
\qquad  \sum_{\omega\in\Omega}\sum_{\alpha\in\Omega^r}n(\alpha\omega)=n$}
\end{eqnarray*}
where finite ``boundary effects",  possibly involving the first or last $r$ symbols of the sequence, are here neglected. Also, noting that  a total of $n(\alpha)!/\prod_\omega n(\alpha\omega)!$ 
permutations of the sequence generate the same $f^D(\omega|\alpha)$, taking the logarithm and using Stirling approximation yields the asymptotic rate formula for  Markov chains
\begin{eqnarray}
 \label{typesformarkov}
 P(f^D|f^M)   \cong   
\exp(- n \: \kappa_{r+1}(f^D||f^M))
\end{eqnarray}
\begin{eqnarray*}
\mbox{\small where}\: 
\mbox{$\displaystyle \kappa_{r+1}(f||g)  :=  K_{r+1}(f||g)-K_{r}(f||g))=
\sum_{\alpha\in \Omega^r} f(\alpha)  \sum_{\omega\in \Omega}f(\omega|\alpha)\: 
\ln \frac{f(\omega|\alpha)}{g(\omega|\alpha)}$}\notag\\
  \mbox{\small and}\qquad 
K_{r}(f||g))   : =   \sum_{\alpha\in \Omega^r}f(\alpha)\: \ln \frac{f(\alpha)}{g(\alpha)}\: .
\notag
\end{eqnarray*}
Setting $r=0$ returns  the asymptotic formula (\ref{typesform}) for independence models. 

\subsection{Testing the order of an empirical sequence}
For $s\le r$, write $\alpha\in \Omega^r$ as $\alpha=(\beta\gamma)$ where
$\beta\in \Omega^{r-s}$ and $\gamma\in \Omega^s$. Consider 
$s$-order models of the form $f^M(\omega|\beta\gamma)=f^M(\omega|\gamma)$. It is not difficult to prove the identity
\begin{eqnarray}
\label{typesformarkov2}
 \min_{f^M\in {\cal  M}_s} \kappa_{r+1}(f^D||f^M)=-H_{r+1}(f^D)+H_{r}(f^D)+H_{s+1}(f^D)-H_{s}(f^D)\notag \\ =
 h_{s+1}(f^D)-h_{r+1}(f^D)\ge0\: .
 \end{eqnarray}
 As an application, consider, as in  section  \ref{tetene} , the log-likelihood nested test of $H_0$  within  $H_1$, opposing  $H_0$ : ``$f^M\in {\cal M}_s$" against 
 $H_1$ : ``$f^M\in {\cal M}_r$". Identities (\ref{typesformarkov}) and  (\ref{typesformarkov2}) lead to the rejection of $H_0$ if
\begin{eqnarray}
\label{rstest}
 2n\:  [h_{s+1}(f^D)-h_{r+1}(f^D)]\ge      
 \chi^2_{1-\alpha}[\scriptstyle{(m-1)(m^r-m^s)}]\: .
 \end{eqnarray} 

 \subsubsection{Example:  test of independence}
For $r=1$ and $s=0$,  the test (\ref{rstest}) amonts in testing independence, and the decision variable
\begin{eqnarray*}
\mbox{\small $h_{1}(f^D)-h_{2}(f^D)=H_{1}(f^D)+H_{1}(f^D)-H_{2}(f^D)=H(X_1)+H(X_2)-H(X_1,X_2)=I(X_1:X_2)$} 
\end{eqnarray*}
is (using stationarity) nothing but the mutual information 
between two consecutive symbols $X_1$ and $X_2$, as expected from example \ref{qqq088}. 

 \subsubsection{Example:  sequential tests}
For $r=1$ and $s=r-1$,  inequality (\ref{rstest}) implies that the model at least of order $r$. Setting $r=1,2,\ldots, r_{\max}$ (with $\mbox{\small df}=(m-1)^2m^{r-1}$) constitutes a sequential procedure  permitting to detect the order of the model, if existing. 

\vspace{0.1in}

For instance, 
a binary Markov chain of order $r=3$ and length $n=1024$ in $\Omega=\{a,b\}$ can be simulated as $X_t:=g(\frac14 (Z_t+Z_{t-1}+Z_{t-2}+Z_{t-3}))$, 
where $Z_t$ are i.i.d. variables uniformly distributed as $\sim U(0,1)$, and
$g(z):=a$ if $z\ge\frac12$ and $g(z):=b$ if $z<\frac12$.  
Application of the procedure at  significance level $\alpha=0.05$ for $r=1,\ldots 5= r_{\max}$ is summarised in the following table, and shows to correctly detect the order of the model: 
\begin{center}\begin{tabular}{|l|l|r|r|r|} \hline
$r$  & $h_r(f^D)$ & $2n[ h_r(f^D)-h_{r+1}(f^D)]$ &   $\mbox{\small df}$ &  $\chi^2_{0.95}[\mbox{\small df}] $   \\ \hline\hline
$1$ & 0.692 &   0.00 &  1 & 3.84\\  \hline
$2$ & 0.692 &   2.05 & 2 & 5.99\\  \hline
$3$ & 0.691 &   {\bf 110.59} & 4 & 9.49\\  \hline
$4$ & 0.637 &   12.29 & 8 & 15.5\\  \hline
$5$ & 0.631 &    18.02 & 16 & 26.3\\  \hline
\end{tabular}
\end{center}

\subsection{Heating and cooling texts}
Let $f(\omega|\alpha)$ (with $\omega\in \Omega$ and $\alpha\in \Omega^r$) denote a conditional distribution of order $r$. In analogy to formula (\ref{bolgib}) of Statistical Mechanics, the distribution can be ``heated" or  ``cooled" at relative temperature $T=1/\beta$ to produce the 
so-called {\em annealed} distribution 
\begin{eqnarray*}
f_\beta(\omega|\alpha):=\frac{f^\beta(\omega|\alpha)}{\sum_{\omega'\in \Omega}f^\beta(\omega'|\alpha)}\: .
\end{eqnarray*}
Sequences generated with the annealed transitions hence simulate texts possessing a  temperature 
$T$ relatively to the original text. 

\subsubsection{Example: simulating hot and cold English texts}
\label{emma}
Conditional distributions of order 3, retaining tetragram structure,  have been calibrated from Jane Austen's  novel {\it Emma}  (1816), containing  $n=868'945$ tokens belonging to  $m=29$ types (the alphabet, the blank, the hyphen and the apostrophe). 
A few annealed simulations are shown below, where the first trigram was sampled from the stationary distribution (Bavaud and Xanthos, 2002). 

\vspace{0.1in}
 
{\bf $\beta=1$} (original process)

{\tt feeliciousnest miss abbon hear jane is arer that isapple did 

ther by 
the withour our the subject relevery that amile 

sament   is laugh in ' 
emma rement on the come februptings he}

\vspace{0.1in}

{\bf $\beta=0.1$} (10 times hotter) 

{\tt torables - hantly elterdays doin said just don't check comedina 

inglas ratefusandinite his happerall bet had had habiticents' 

oh young most brothey lostled wife favoicel let
you cology}

\vspace{0.1in}

{\bf $\beta=0.01$} (100 times hotter): any transition having occurred in the original text tends to occur again with uniform probability, making the heated text maximally unpredictable.
However, most of the
possible transitions did not  occur initially, which explains the persistence
of the English-like aspect. 

\vspace*{0.1cm}

{\tt et-chaist-temseliving dwelf-ash  eignansgranquick-gatefullied  

georgo namissedeed fessnee th   thusestnessful-timencurves -

 him 
duraguesdaird vulgentroneousedatied yelaps isagacity in}

\vspace{0.1in}

{\bf $\beta=2$} (2 times cooler) : conversely, frequent (rare) transitions become even more frequent (rare), making the text fairly predictable.
 
\vspace*{0.1cm}

{\tt 's good of his compassure is a miss she was she come to the

 of his 
and as it it was so look of it i do not you with her

 that i  am
superior the in ther which of that the half - and}

\vspace{0.1in}

{\bf $\beta=4$} (4 times cooler): in the low temperature limit, dynamics is trapped in the most probable initial transitions and texts properly become crystal-like, as expected from Physics (see example
 \ref{qqq444}):

\vspace*{0.1cm}

{\tt ll the was the was the was the was the was the was the was

 the
was the was the was the was the was the was the was the

 was the  
was  the   was the was the was the was the was the was}

\subsection{Additive and multiplicative text mixtures}
 In the spirit of section \ref{lefd}, {\em additive} and  {\em multiplicative} 
 mixtures of 
 two conditional distributions
$f(\omega|\alpha)$ and $g(\omega|\alpha)$ of order $r$ can be constructed as
\begin{equation*}
\mbox{\small $h_{\lambda}(\omega|\alpha):=\lambda f(\omega|\alpha)+(1-\lambda)g(\omega|\alpha)\qquad \quad
h_{\mu}(\omega|\alpha):=\frac{f^{\mu}(\omega|\alpha)\:  g^{(1-\mu)}(\omega|\alpha)}
{\sum_{\omega'\in \Omega}f^{\mu}(\omega'|\alpha)\:  g^{(1-\mu)}(\omega'|\alpha)}$}
\end{equation*}
where $0<\lambda<1$ and $0<\mu<1$. The resulting transition exists if it exists in at least one of the initial distributions (additive mixtures) or in both distributions (multiplicative mixtures). 

\subsubsection{Example: additive mixture of English and French}
\label{linkasjdh}
Let $g$ denote the empirical distribution of order 3 of example (\ref{emma}), and define $f$ as the corresponding  distribution estimated on the $n=725'001$ first symbols  of the French novel {\em La b\^{e}te humaine} from Emile Zola. Additive simulations with various values of $\lambda$ read  (Bavaud and Xanthos, 2002): 

\vspace{0.1in}
 
{\bf $\lambda=0.17$}  

{\tt ll thin not alarly but alabouthould only to comethey had be

 the sepant a was que lify you i  
bed at it see othe to had 

state cetter but of i she done a la veil la preckone forma feel} 

\vspace{0.1in}
 
{\bf $\lambda=0.5$}  

{\tt daband shous ne findissouservait de sais comment do be certant

 she cette l'ideed se point le 
fair somethen l'autres jeune suit 

onze muchait satite a ponded was si je lui love toura}

\vspace{0.1in}
 
{\bf $\lambda=0.83$}

{\tt les appelleur voice the toodhould son as or que aprennel un

 revincontait en at on du semblait juge yeux 
plait etait 

resoinsittairl on in and my she comme elle ecreta-t-il avait 

autes foiser}

\vspace{0.1in}

showing, as expected, a gradual transformation from   English-  to French-likeness 
with increasing $\lambda$. 

\subsubsection{Example: multiplicative mixture of English and French}
Applied now on multiplicative mixtures, 
the procedure described in example \ref{linkasjdh}  yields  (Bavaud and Xanthos, 2002)

\vspace{0.1in}
 
{\bf $\mu=0.17$}  

{\tt licatellence a promine agement ano ton becol car emm*** ever 

ans touche-***i harriager gonistain ans tole elegards intellan 

enour
bellion genea***he succept wa***n instand instilliaristinutes}

\vspace{0.1in}
 
{\bf $\mu=0.5$}  

{\tt n neignit innerable quit tole ballassure cause on an une grite  

chambe ner martient infine disable prisages creat
mellesselles 

dut***grange accour les norance trop mise une les emm***}

 \vspace{0.1in}
 
{\bf $\mu=0.83$}  

{\tt es terine fille son mainternistonsidenter ing sile celles 

tout a pard elevant poingerent une graver dant lesses 

jam***core
son luxu***que eles visagemensation lame cendance}

\vspace{0.1in}

where the symbol {\tt ***} indicates that the process is trapped in a trigram occuring in the  English, but not in the French sample (or vice versa). Again, the French-likeness of the texts increases with $\mu$. Interestingly enough, some simulated subsequences are arguably  evocative of Latin, whose lexicon contains an important part of the forms common to 
English and French.

\vspace{0.1in}

From an inferential point of view, the   multiplicative mixture is of the form (\ref{neopea}), and hence
lies at the boundary of the optimal Neyman-Pearson decision region, governing the asymptotic rate of errors of both kinds, namely confounding  French with English or English with French.

\section{Bibliography}
 \begin{itemize}
\item[$\bullet$] Amari, S.-I.  {\em Differential-Geometrical Methods in Statistics}, Lecture Notes in Statistics {\bf 28}, Springer (1985)
\item[$\bullet$] Bavaud, F. {\em The Quasisymmetric Side of Gravity Modelling}, Environment and Planning A, 34, pp.61-79 (2002a) 
\item[$\bullet$] Bavaud, F. {\em Quotient Dissimilarities, Euclidean Embeddability, and Huygens's Weak Principle}, in
{\em Classification, Clustering and Data Analysis} Jajuga, K., Solkolowski, A.  and Bock, H.-H. (Eds.), pp.195-202,  Springer (2002b)
\item[$\bullet$] Bavaud, F. and Xanthos, A. {\em  Thermodynamique  et Statistique Textuelle: concepts  et illustrations}, in
{\em Proceedings of JADT 2002}  (6\`emes Journ\'ees internationales d'Analyse statistique des 
Donn\'ees Textuelles), St-Malo, (2002)
\item[$\bullet$] Billingsley, P. {\em Statistical Inference for Markov Processes},
University of Chicago Press, Chicago  (1961)
\item[$\bullet$]  Bishop, Y.M.M., Fienberg, S.E. and Holland, P.W. {\em Discrete multivariate Analysis}, 
The MIT Press, Cambridge, (1975) 
\item[$\bullet$] Boltzmann, L. {\em Weitere Studien \"{u}ber das W\"{a}rmegleichgewicht unter 
Gasmolek\"{u}len}, Sitzungsberichte
der Akademie der Wissenschaften {\bf 66} pp.275-370 (1872) 
\item[$\bullet$] Cardoso, J.-F. {\em Dependence, Correlation and Gaussianity in Independent Component Analysis}, Journal of Machine Learning Research {\bf 4} pp.1177-1203 (2003)
\item[$\bullet$]  Caussinus, H.  {\em  Contribution \`a l'analyse statistique des tableaux de corr\'e-lation}, Annales de la Facult\'e des Sciences de Toulouse {\bf 29}   pp.77-183  (1966)
\item[$\bullet$]  Christensen, R.   {\em Log-Linear Models}, Springer (1990)
\item[$\bullet$] Cover, T.M. and Thomas, J.A. {\em Elements of Information Theory}, Wiley (1991)
\item[$\bullet$] Cramer, H. {\em Mathematical Methods of Statistics}, Princeton University Press (1946)
\item[$\bullet$] Csisz{\'a}r, I. {\em $I$-Divergence Geometry of Probability Distribution and Minimization Problems}, The Annals of Probability {\bf 3}, pp.146-158 (1975)
\item[$\bullet$]  Csisz{\'a}r, I.  and K\"{o}rner, J. {\em Towards a general theory of source networks}, IEEE Trans. Inform. Theory {\bf 26}, pp.155-165 (1980)
\item[$\bullet$]  Csisz{\'a}r, I.  and Tusn{\'a}dy, G. {\em Information Geometry and Aternating Minimization Procedures}, Statistics and Decisions, Supplement Issue {\bf 1}  pp.205-237 (1984)
\item[$\bullet$]  Dempster, A.P., Laird, N.M and Rubin, D.B.  {\em Maximum Likelihood from Incomplete Data via the EM Algorithm}, J. Roy. Stat. Soc. B  {\bf
39} pp.1-22   (1977)
\item[$\bullet$]  Ferguson, T.S. {\em Prior Distributions on Spaces of Probability Measures}, The Annals of Statistics {\bf 2} pp.615-629   (1974)
\item[$\bullet$]  Jaynes, E.T. {\em Information theory and statistical mechanics}, Physical Review {\bf
106} pp.620-630 and {\bf 108} pp.171-190 (1957)
\item[$\bullet$] Jaynes, E.T. {\em Where do we stand on maximum entropy?}, presented at the Maximum Entropy Formalism Conference, MIT, May 2-4 (1978)
\item[$\bullet$] Kullback, S. {\em Information Theory and Statistics}, Wiley (1959)
\item[$\bullet$]  Lee, T.-W., Girolami, M., Bell, A.J. and Sejnowski, T.J. {\em A unifying
 Information-Theoretic Framework for Independent Component Analysis}, Computers and Mathematics with Applications  {\bf 39}  pp.1-21 (2000)
\item[$\bullet$] Li, M. and Vitanyi, P.  {\em An Introduction to Kolmogorov
complexity and its applications}, Springer (1997)
\item[$\bullet$] MacKay, D.J.C. {\em Information Theory, Inference and Learning Algorithms}, Cambridge University Press (2003)
\item[$\bullet$]  Popper, K. {\em Conjectures and Refutations}, Routledge (1963)
\item[$\bullet$] Robert, C.P. {\em The Bayesian Choice}, second edition, Springer (2001)  
\item[$\bullet$] Sanov, I.N. {\em On the probability of large deviations of random variables}, Mat. Sbornik  {\bf 42} pp.11-44   (1957) (in Russian. English translation in Sel. Trans. Math. Statist. Probab.  {\bf 1} pp.213-244   (1961))
\item[$\bullet$] Saporta, G. {\em Probabilit\'es, Analyse de Donn\'ees et Statistique}, Editions Technip, Paris, (1990)
\item[$\bullet$] Simon, G. {\em Additivity of Information in Exponential Family Power Laws}, Journal of the American Statistical Association, {\bf 68}, pp.478-482 (1973)
\item[$\bullet$] Shannon, C.E. {\em A mathematical theory of communication}, Bell System Tech. J. {\bf 27}, pp.379-423; 623-656 (1948)
\item[$\bullet$]  Tribus, M. and McIrvine, E.C. {\em Energy and Information}, Scientific American {\bf 224} pp 178-184 (1971)
\item[$\bullet$] Vapnik, V.N. {\em The Nature of Statistical Learning Theory}, Springer, (1995)
\end{itemize}

\end{document}